\documentclass[twocolumn,showpacs,preprintnumbers,amsmath,amssymb,superscriptaddress]{revtex4}

\usepackage{graphicx}
\usepackage{dcolumn}
\usepackage{bm}
\usepackage{amsmath}
\usepackage{amssymb}
\usepackage{latexsym}
\usepackage{setspace}
\usepackage{graphics}
\usepackage{amsfonts}
\usepackage{epsfig}
\usepackage{enumerate}
\def\lsim{\mathrel{\rlap{\lower4pt\hbox{\hskip1pt$\sim$}}
    \raise1pt\hbox{$<$}}}         
\def\gsim{\mathrel{\rlap{\lower4pt\hbox{\hskip1pt$\sim$}}
    \raise1pt\hbox{$>$}}}         

\graphicspath{%
    {converted_graphics/}
    {/}
}
\begin{document}

\preprint{APS/123-QED}

\title{Second harmonic light scattering induced by defects in the twist-bend nematic phase of achiral liquid crystal dimers\\}

\author{Shokir A. Pardaev}
\affiliation{%
Department of Physics, Kent State University, Kent, Ohio 44242, USA\\
}%

\author{S. M. Shamid}
\affiliation{%
Department of Physics, Kent State University, Kent, Ohio 44242, USA\\
}%


\author{M. G. Tamba}
\affiliation{%
Department of Chemistry, University of Hull, Hull HU6 7RX, UK\\
}%
\author{C. Welch}
\affiliation{%
Department of Chemistry, University of Hull, Hull HU6 7RX, UK\\
}%

\author{G. H. Mehl}
\affiliation{%
Department of Chemistry, University of Hull, Hull HU6 7RX, UK\\
}%

\author{J. T. Gleeson}
\affiliation{%
Department of Physics, Kent State University, Kent, Ohio 44242, USA\\
}%

\author{D. W. Allender}
\affiliation{%
Department of Physics, Kent State University, Kent, Ohio 44242, USA\\
}%

\author{J. V. Selinger}
\affiliation{%
Chemical Physics Interdisciplinary Program and Liquid Crystal Institute, Kent State University, Kent, Ohio 44242, USA\\
}%

\author{B. Ellman}
\affiliation{%
Department of Physics, Kent State University, Kent, Ohio 44242, USA\\
}%

\author{A. Jakli}
\affiliation{%
Chemical Physics Interdisciplinary Program and Liquid Crystal Institute, Kent State University, Kent, Ohio 44242, USA\\
}%

\author{S. Sprunt}
\email{ssprunt@kent.edu}
\affiliation{%
Department of Physics, Kent State University, Kent, Ohio 44242, USA\\
}%

\date{\today}

\begin{abstract}
The nematic twist-bend ($\mathrm{N_{TB}}$) phase, exhibited by certain thermotropic liquid crystalline (LC) dimers, represents a new orientationally ordered mesophase -- the first distinct nematic variant discovered in many years. The $\mathrm{N_{TB}}$ phase is distinguished by a heliconical winding of the average molecular long axis (director) with a remarkably short (nanoscale) pitch and, in systems of achiral dimers, with an equal probability to form right- and left-handed domains. The $\mathrm{N_{TB}}$ structure thus provides another fascinating example of spontaneous chiral symmetry breaking in nature. The order parameter driving the formation of the heliconical state has been theoretically conjectured to be a polarization field, deriving from the bent conformation of the dimers, that rotates helically with the same nanoscale pitch as the director field. It therefore presents a significant challenge for experimental detection. Here we report a second harmonic light scattering (SHLS) study on two achiral, $\mathrm{N_{TB}}$-forming LCs, which is sensitive to the polarization field due to micron-scale distortion of the helical structure associated with naturally-occurring textural defects. These defects are parabolic focal conics of smectic-like ``pseudo-layers", defined by planes of equivalent phase in a coarse-grained description of the $\mathrm{N_{TB}}$ state. Our SHLS data are explained by a coarse-grained free energy density that combines a Landau-deGennes expansion of the polarization field, the elastic energy of a nematic, and a linear coupling between the two.
\end{abstract}


\maketitle

\section{Introduction}
Polar orientational order in liquid crystals (LCs) is traditionally associated with the hindered rotation of chiral molecules arranged in layers (smectic phase). In conventional ferroelectric LCs, this hindrance is caused by tilt of molecules with respect to the layer normal (smectic-C phase), which leads to broken inversion symmetry and a spontaneous polarization parallel to the layers. The development and exploration in the past 20 years of mesogens with a bent-shaped (rather than rod-like) rigid core structure has seen the requirement of molecular chirality eliminated: In these systems of achiral molecules, both smectic-C and smectic-A (untilted) phases exhibit in-layer polarization and, perhaps more interestingly, in the smectic-C case spontaneously separate into domains of opposite structural chirality. Layer tilt plus layer polarization therefore induce chirality.

So far, however, the realization of polar orientational order {\it without} layering or constituent chirality has proven elusive. While simple nematics exhibit a typically weak polarization under an applied orientational stress (the so-called flexoelectric effect \cite{Meyer_flexo}), in equilibrium the magnitude of the vector order parameter representing the dipole moment of the molecules vanishes. The recent discovery of the twist-bend nematic ($\mathrm{N_{TB}}$) phase \cite{Cestari_PRE,Borshch_Nature,Chen_PNAS}, originally predicted by Meyer \cite{Meyer_TB}, may have fundamentally altered this situation. 

The $\mathrm{N_{TB}}$ phase typically occurs at temperatures below the ordinary uniaxial nematic phase, in achiral liquid crystal dimers composed of a pair of rodlike mesogenic units connected by an odd-numbered hydrocarbon linkage. This linkage favors an overall bent conformation (Fig.~1(a)). As shown in Fig.~1(b), the $\mathrm{N_{TB}}$ state is characterized by a modulated orientational structure in which the average molecular long axis (local director $\hat{\mathbf{n}}$) simultaneously bends and twists in space in a periodic fashion, producing a heliconical configuration. The typical cone angle (average tilt of the molecules away from the helical axis, inferred from optical birefringence measurements) \cite{Borshch_Nature,CMeyer3}, is $\beta \sim 10^\circ$. The pitch ($t_0$) of the modulation, measured directly by freeze-fracture TEM \cite{Chen_PNAS}, is surprisingly short -- of order a few molecular lengths (i.e., $t_0 \sim 10$~nm). Both parameters differ markedly from the usual chiral nematic (cholesteric), where $\beta = 90^\circ$ and $t_0 \gtrsim 100$~nm.

Various theoretical models \cite{Dozov_TB,CMeyer2,Shamid_PRE,Kats,Virga,Greco,Barbero,Lelidis,Vanakaras} have been proposed to account for the properties of the $\mathrm{N_{TB}}$ phase and the nature of the nematic-$\mathrm{N_{TB}}$ transition. Dozov \cite{Dozov_TB} and Meyer et al \cite{CMeyer2} developed a model based on the Frank elasticity for bend distortions of $\hat{\mathbf{n}}$ becoming negative at the transition; in this thoery, the order parameter for the $\mathrm{N_{TB}}$ state is $\sin^2\beta$. Two other models -- those of Kats and Lebedev \cite{Kats} and Shamid et al \cite{Shamid_PRE} -- introduce a vector order parameter and expand the Landau-deGennes free energy density in terms of this order parameter and $\hat{\mathbf{n}}$.  In particular, Shamid et al \cite{Shamid_PRE} propose that the order parameter is a helical polarization wave, $\mathbf{P}$, which is orthogonal to, but has the same nanoscale pitch as, the heliconical director $\hat{\mathbf{n}}$. ($\mathbf{P}$ may represent a shape polarization in lieu of, or in addition to, an electric polarization.) A linear coupling of $\mathbf{P}$ to the curl of $\hat{\mathbf{n}}$ drives the heliconical modulation of $\hat{\mathbf{n}}$, and, on the nematic side of the transition, renormalizes the bend elastic constant to zero, destabilizing the nematic phase in agreement with the theory of Dozov \cite{Dozov_TB} and with experiment \cite{Adlem}. While studies of fluctuations in the $\mathrm{N_{TB}}$ phase are consistent with the presence of a short-pitch polarization wave \cite{Parsouzi_PRX}, more direct confirmation, by a method sensitive to broken centrosymmetry due to $\mathbf{P}$, is clearly desirable. Given the absence of any mass density wave (Bragg peaks) in X-ray diffraction measurements \cite{Chen_PNAS,Mehl_M37,Ahmed}, confirmation of $\mathbf{P}$ would establish the $\mathrm{N_{TB}}$ state as the first example of a locally polar LC phase without smectic-like molecular layering -- a prospect of fundamental significance. 

This paper presents the results of a second harmonic light scattering (SHLS) study designed to test this prospect. In order to detect at optical wavelengths a polarization field that averages out over a nanoscale helical pitch, our approach takes advantage of another interesting feature of the $\mathrm{N_{TB}}$ phase -- the presence of topological defects, analogous to the classical parabolic focal conic (PFC) defects in the layer structure of simple smectic LCs, that distort the polarization field over a length scale much larger than $t_0$ and thereby produce a nonzero $\mathbf{P}$ on the optical scale suited to the SHLS technique \cite{Pardaev_PRE}. 

\begin{figure}[t]
\centering
\includegraphics[width=1\columnwidth]{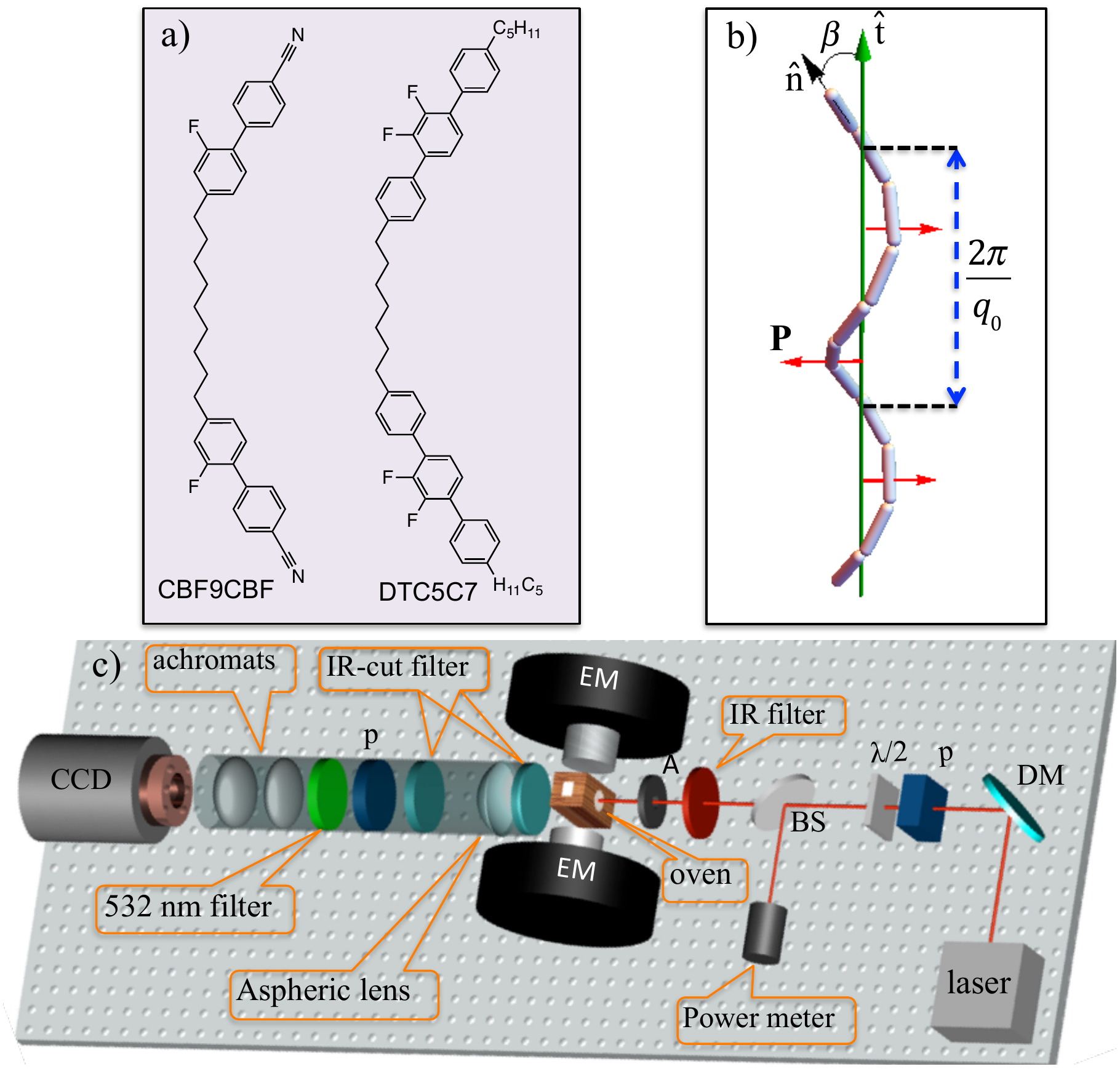}
\caption{(a): Structure of LC dimers that form the $\mathrm{N_{TB}}$ phase in the mixture KA(0.2) and the single compound DTC5C7. (b): Schematic rendering of the heliconical molecular orientation in the $\mathrm{N_{TB}}$ phase, with cone angle $\beta$ and helical wavenumber $q_0$. Each cylinder represents the average molecular long axis of a dimer, and the vectors $\hat{\mathbf{n}}$ and $\mathbf{P}$ correspond to the local director and polarization fields, respectively, while $\hat{\mathbf{t}}$ is the average or ``coarse-grained" director. (c): 3D CAD rendering of the experimental apparatus.}
\end{figure}

Specifically, we demonstrate an SHLS signal in the $\mathrm{N_{TB}}$ phase, whose key features (optical polarization selection rules, in particular) are explained by a model in which slabs of the $\mathrm{N_{TB}}$ system, defined by planes of constant helical phase, are treated as ``pseudo-layers". When their spacing is reduced due to a temperature dependent helical pitch, these ``layers" exhibit PFC defects in an a analogous manner as true smectics do when placed under a dilatory strain. The concept of ``pseudo-layers" was previously introduced in the development of a ``coarse-grain" theory of cholesterics \cite{deGennes,Lubensky_PRA,Radzihovsky_PRE}. 

This concept has recently been applied to the $\mathrm{N_{TB}}$ phase to describe smectic-like elastic properties \cite{CMeyer3,Parsouzi_PRX} and fluctuation modes \cite{Parsouzi_PRX}, strong shear thinning behavior \cite{Jakli1}, and optical stripe and focal textures \cite{Jakli2}. Starting from a theoretically proposed free energy density coupling $\hat{\mathbf{n}}$ and $\mathbf{P}$ \cite{Shamid_PRE}, we show that ``pseudo-layer" deformations associated with PFCs naturally lead to a distortion of the helical field $\mathbf{P}$ that breaks centrosymmetry at length scales comparable to (or greater than) optical wavelengths. We also verify that the SHLS signal is absent in an ordinary calamitic smectic liquid crystal containing a comparable density of PFCs, but lacking any underlying polar structure. Our results and analysis thus provide strong evidence of an underlying helical polarization field that characterizes the $\mathrm{N_{TB}}$ phase.

\section{Experimental Details}

The $\mathrm{N_{TB}}$ materials studied are: (1) a mixture [denoted KA(0.2)] \cite{Adlem} of five odd-membered dimers with ether linkages between mesogenic groups plus one dimer with a methylene linkage, CBF9CBF (Fig.~1(a)), which is considered to be the active component in inducing the $\mathrm{N_{TB}}$ phase; and (2) a pure dimer \cite{Mehl_M37,Tamba} (labeled DTC5C7 and also shown in Fig.~1(a)). The phase sequences of these materials (in cooling) are isotropic-($77^\circ$)-N-($37.4^\circ$)-$\mathrm{N_{TB}}$-($22^\circ$C)-crystal [KA(0.2)] and isotropic-($156.8^\circ$)-N-($127.8^\circ$)-$\mathrm{N_{TB}}$-($96.6^\circ$)-$\mathrm{SmX}$-($77.4^\circ$C)-crystal (DTC5C7). As a control compound, we chose the conventional, rodlike thermotropic, 4-n-octyl-4'-cyanobiphenyl (8CB).

The liquid crystals were either capillary filled into sandwich cells made of fused silica (FS) substrates with a 0.1~mm spacer or loaded into FS cuvettes with a 1~mm path length. In both cases the substrate surfaces were thoroughly cleaned but otherwise untreated. The loaded cells were placed in a temperature-regulated oven with optical access on opposite sides and with $\sim 0.01$~K stability. The oven fitted between the pole faces of a laboratory electromagnet, with the direction of the field oriented perpendicular to the substrate normal (Fig.~1(c)). The director $\hat{\mathbf{n}}$ in the nematic phase or direction $\hat{\mathbf{t}}$, corresponding to the average of $\hat{\mathbf{n}}$ over the heliconical pitch in the $\mathrm{N_{TB}}$ phase (see Fig.~1(b)), was magnetically aligned by slowly cooling from the isotropic phase in a 1.13~T field produced by the magnet.

The samples were illuminated at normal incidence to the substrates and to the direction of $\hat{\mathbf{t}}$ with polarized 1064 nm light from a Nd:YAG laser (Continuum Minilite II, outputting 5-7~ns, $\sim 2$~mJ pulses at 10~Hz). The beam diameter at the sample was reduced to 1~mm by a fixed aperture. At the incident pulse energy used, no evidence of damage to the samples was detected during or after the measurements. The incident polarization could be rotated between vertical ($V$, perpendicular to $\hat{\mathbf{t}}$) and horizontal ($H$, parallel to $\hat{\mathbf{t}}$) orientations. The directions $V$, $H$ also refer to vertical and horizontal with respect to the plane of the apparatus in Fig.~1(c).

The scattered second harmonic (SH) light from the samples was collected in a cone of $15^\circ$ opening angle around the forward (transmitted light) direction by a specially constructed optical telescope, which imaged the scattering pattern onto the sensor of a cooled CCD camera (Princeton Instruments, model ProEM512). An analyzer was used to set the $V$ and $H$ polarization state of the SH light, while a series of broad and narrow band filters (the latter having a passband of 1~nm FWHM at 532~nm) removed the transmitted fundamental light. 

The telescope could be interchanged with a polarizing optical microscope for in-situ examination of the sample textures over the region illuminated by the laser beam and for confirmation of uniform director alignment. SH scattering patterns and optical micrographs were recorded for various polarizer/analyzer combinations at fixed temperatures in the $\mathrm{N_{TB}}$ phase of the samples. 

\begin{figure}[t]
\centering
\includegraphics[width=1.0\columnwidth]{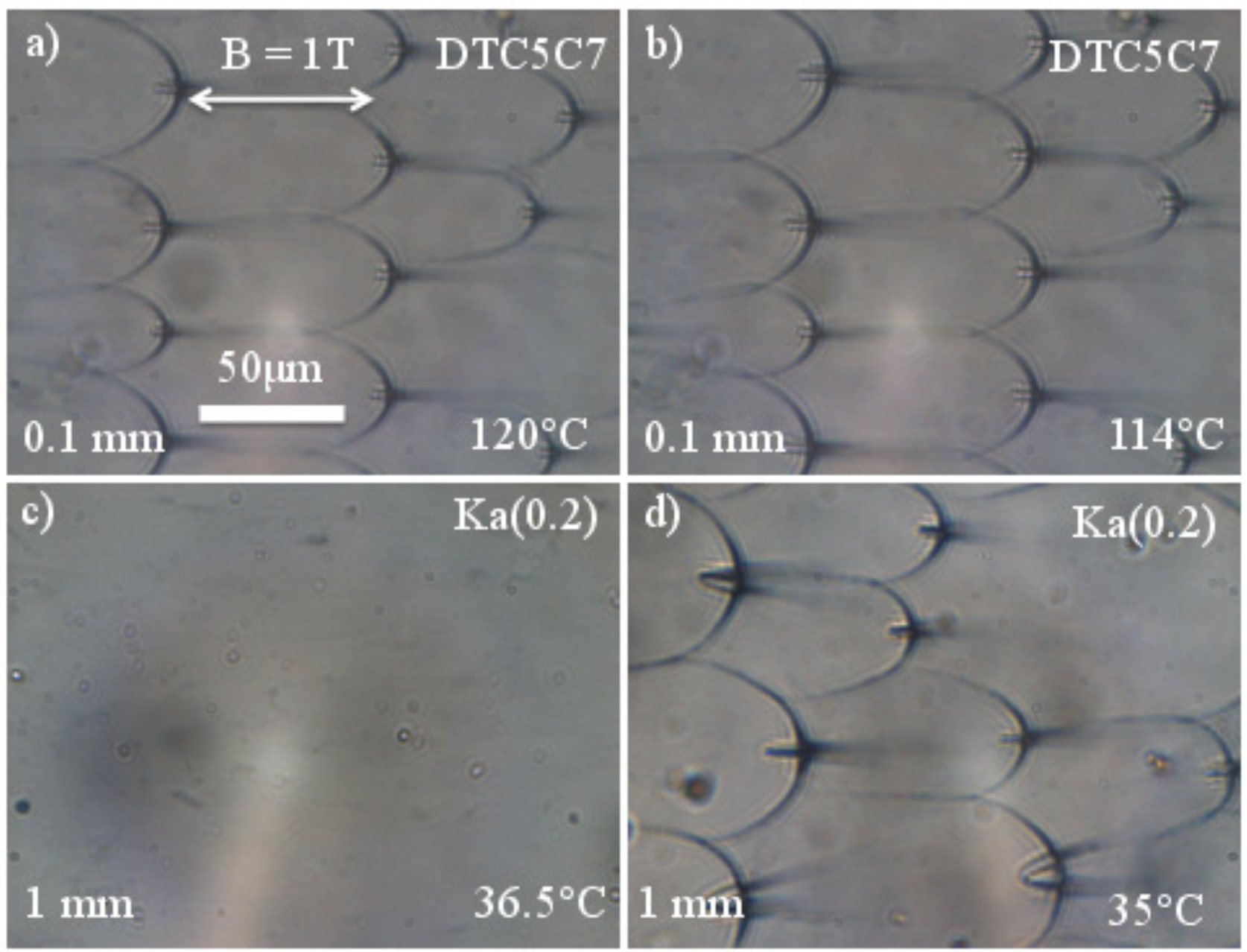}
\caption{Polarizing microscope images of parabolic focal conic defect arrays in the $\mathrm{N_{TB}}$ phase of the studied LC compounds (polarizer along horizontal direction, analyzer at $10^\circ$ angle from horizontal). (a) and (b): Images taken in the mid-plane of a 0.1 mm thick sample of DTC5C7 at temperatures $7.8^\circ$C and $13.8^\circ$C below the unixial nematic to $\mathrm{N_{TB}}$ transition. (These temperatures correspond to two of the data sets for DTC5C7 in Fig.~3.) (c): Image taken near the surface of a 1~mm cell of KA(0.2), $0.9^\circ$C below the transition. There are no observable PFCs and no SH signal (see Fig.~3). (d): At $\sim 2^\circ$C below the transition, PFCs appear and an SH signal is observed. All of the images were recorded after cooling through the transition in a 1.13~T magnetic field (horizontal in the figure).}
\end{figure}

\begin{figure*}[tbp]
\centering
\includegraphics[width=2.05\columnwidth]{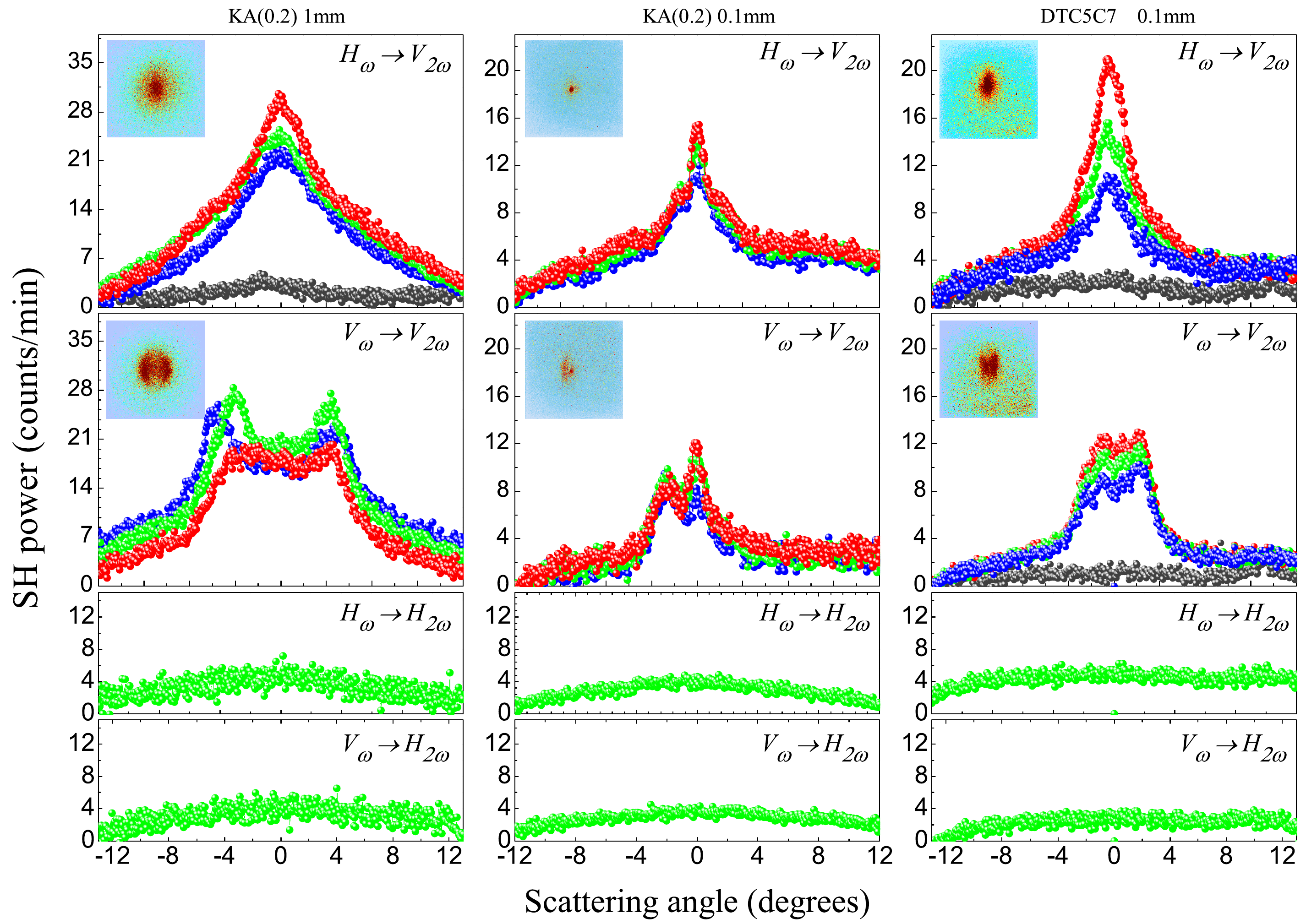}
\caption{Angular distribution of SH light recorded in the $\mathrm{N_{TB}}$ phase of KA(0.2) and DTC5C7. The thumbnail images (insets) show the raw data from the CCD detector, color coded for intensity (light blue representing the background level, dark red representing peak level). The axis of the applied magnetic field (and average LC director) is horizontal in the thumbnails. The main figures display plots SH power vs scattering angle for a horizontal cut (40 pixels wide) through the central peak. The polarizer and analyzer combinations for the normally incident fundamental light (wavevector $\mathbf{k}_\omega$) and the second harmonic light collected from the sample are indicated in each panel; $H$ (horizontal) corresponds to polarization parallel to the average LC director $\hat{\mathbf{t}}$ and also to the axis through the foci of the parabolae delineating the PFC defects in Fig.~2, while $V$ (vertical) applies to the direction perpendicular to $\hat{\mathbf{t}}$ and to $\mathbf{k}_\omega$. First two columns (from left): Data for KA(0.2) in thick (1~mm) and thin (0.1~mm) optical cells: Note the approximately twofold decrease in signal, despite a tenfold decrease in sample thickness. Data, taken in cooling, are shown for several temperatures relative to the nematic to $\mathrm{N_{TB}}$ transition: $-0.9^\circ$C (gray), $-3.4^\circ$C (blue), $-5.4^\circ$C (green), and $-7.4^\circ$C (red points). For the temperatures close to the transition, no PFCs are observed (Fig.~2), and there is no SH signal. Note that the central peak is split when the incident (fundamental light) polarization is switched from $H$ to $V$ (for $V$ SH output), but there is no $H$ output for either $H$ or $V$ input polarization (bottom panels). Third column: Similar results for a thin cell containing DTC5C7, at temperatures $-3.8^\circ$C (gray), $-7.8^\circ$C (blue), $-9.8^\circ$C (green), and $-13.8^\circ$C (red) relative to the nematic to NTB transition. Again there is no $H$ polarized SH output.}
\end{figure*}

\section{Results}

\subsection{Microscopy}

First we consider the results from polarizing microscopy displayed in Fig.~2; the images shown were obtained after cooling the samples in the applied magnetic field to temperatures below the transition temperature ($T_{TB}$) to the $\mathrm{N_{TB}}$ phase. In the images, the polarizer and analyzer axes are approximately parallel to the field. The effective director $\hat{\mathbf{t}}$ is uniformly oriented along the field, except in the immediate vicinity of parabolic focal conic (PFC) defects. These are delineated in Fig.~2 by dark lines tracing out parabolic trajectories, which are paired together in orthogonal planes such that each parabola in a pair passes through the other's focus. The paired parabolae resemble the classic PFC texture observed in ordinary smectic LCs under a dilatory strain of the smectic layers \cite{Rosenblatt_PFC}.

In the images of Fig.~2, typically one parabola lies in the focal plane (parallel to the cell substrates), while the second of the pair occupies a plane normal to the viewing direction. The spacing between the two opposite ``arms" within each PFC is $\sim 50 \mu$m, while the separation between foci in the core region is about $6~\mu$m. The symmetry axes of the PFCs are well-aligned, and parallel to the $\mathrm{N_{TB}}$ optical axis ($\hat{\mathbf{t}}$) direction. In the thicker (1~mm) cells, the core regions of the PFCs could be resolved in two distinct planes close to the opposing cell boundaries, whereas in the 0.1~mm cells a single focal plane was located at essentially the mid-plane of the cell. Thus, two layers of PFCs form near the opposing substrates in sufficiently thick samples, but are squeezed into a single layer when the cell thickness becomes comparable to the dimensions of the parabolae.

As Fig.~2(c) reveals, the PFCs are absent at temperatures close to (within $\sim 1^\circ$ of $T_{TB}$); they appear and their population increases in the range $T_{TB}-T \simeq 1$ to $3^\circ$C, and then saturates. By analogy with ordinary smectics, we interpret the PFCs as defects in a ``pseudo-layer" structure defined by $\sim 10$~nm thick slabs between planes of constant phase in the heliconical $\mathrm{N_{TB}}$ structure. As the temperature decreases below $T_{TB}$, the ``layer" spacing (pitch $t_0$) shrinks, resulting in PFC formation at a threshold $t_0$ just as a critical dilatory strain does in the case of true smectic layers. 

\subsection{Second harmonic light scattering}

We now turn to our results from SHLS measurements in the $\mathrm{N_{TB}}$ phase. These are summarized in Fig.~3. First consider the case where the incident (fundamental) light is polarized along $\hat{\mathbf{t}}$ and the SH output is polarized perpendicular to $\hat{\mathbf{t}}$ (an $H_\omega \rightarrow V_{2\omega}$ process, following the nomenclature introduced above). In both $\mathrm{N_{TB}}$ materials, for $T$ below but within $1^\circ$C of the transition ($T_{TB}$) -- the range where PFCs are absent -- no SH output is detected. Below this range and in the presence of PFCs, a clear SH signal is observed, with a peak in the forward direction. In the KA(0.2) samples, the signal level increases slightly with decreasing $T$ in both thick and thin samples, whereas the signal approximately doubles for DTC5C7. The angular FWHM of the peak (using data from the thin cell where one expects less broadening due to multiple scattering) corresponds to a length scale of $1.22 \lambda_{2\omega}/$FWHM~$\sim 7~\mu$m, which is comparable to the spacing between foci of the PFCs. The SH signal is clearly associated with the appearance of PFCs in the $\mathrm{N_{TB}}$ phase.

For $V$ polarized input and $V$ output ($V_\omega \rightarrow V_{2\omega}$ process), the angular distribution of SH light is again concentrated around the forward direction. However, the peak is split approximately symmetrically about zero scattering angle (compare the thumbnail images in Fig.~3 showing the raw SHLS patterns recorded on the CCD). Also, the temperature dependence is  weaker than for the $H_\omega \rightarrow V_{2\omega}$ process. On the other hand, for $H$ polarized output ($H_\omega \rightarrow H_{2\omega}$ and $V_\omega \rightarrow H_{2\omega}$ processes), there is {\it no} SH signal. Thus, {\it all detected SH output is polarized perpendicular to the average helical axis of the TB structure or, equivalently, to the axis of the PFC defects}.

The peak SH intensity from the 0.1~mm thick samples of KA(0.2) is approximately two times lower than in the 1~mm sample. For incoherent SH generation from the bulk NTB phase, we would normally expect a factor $\sim 10$ decrease. The discrepancy is explained by the SHG originating from the PFC defects: As noted above, there are two ``layers" of PFCs, localized near the surfaces, in the thick cell, but only one in the thin cell -- hence the factor of $\sim 2$ change observed in SH intensity.

In all cases, the SH signal is quite weak -- of order $\sim 10^{-7}$ the level from a 0.5~mm thick Z-cut single crystal quartz reference. Several factors, beyond the possibility of an intrinsically small nonlinear susceptibility, may contribute to this weak level. First, the scattering is incoherent, and there is no phase matching between the fundamental and second harmonic waves. Second, the signal comes only from the PFCs, and the region of strong ``pseudo-layer" deformation (defined by the parabolic lines in the images in Fig.~2) occupies only a small fraction of the total volume illuminated by the fundamental laser beam. Third, as we shall describe later in the Discussion section, cancellations in the induced polarization (and effective non-centrosymmetry) occur both at points on the parabolae and between points on opposite sides of them, reducing the SH output.

To confirm that the SH signal arises specifically from deformation of $\mathrm{N_{TB}}$ ``pseudo-layers", we performed similar measurements on the smectic-A phase of an ordinary rod-like LC, octyl cyanobiphenyl (8CB), which also exhibits classical PFCs. As shown in Fig.~4, for both 1~mm and 0.1~mm cells populated with PFC defects of an ordinary smectic layer structure, there is no detectable SH signal under experimental conditions identical to those used for the $\mathrm{N_{TB}}$ samples.

\begin{figure}[t]
\centering
\includegraphics[width=1.0\columnwidth]{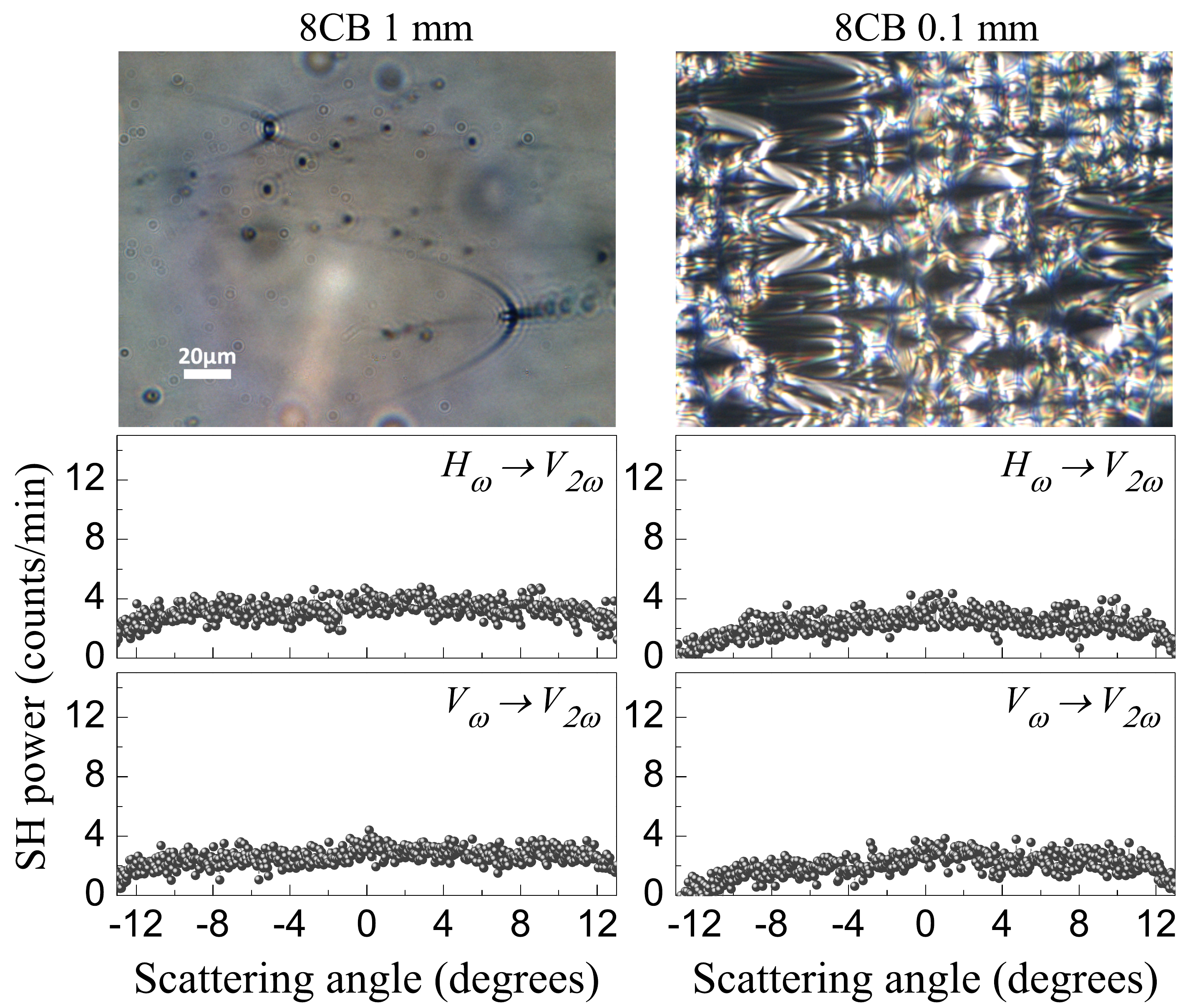}
\caption{Results for 1 and 0.1~mm thick cells of the control sample of 8CB in the smectic-A phase. No SH output is observed despite the presence of PFCs in the smectic layer structure (see accompanying microscope images).}
\end{figure}

\section{Discussion}

Let us now consider mechanisms for second harmonic generation due to the presence of PFCs, which could potentially account for the experimental results described in the previous section. Before beginning, it may be useful to summarize the following key points of the analysis developed in the subsections below:
\begin{enumerate}[{(1)}]
\item Deformation of the $\mathrm{N_{TB}}$ ``pseudo-layers"  and the average director, due to PFCs, distorts the helical polarization field associated with $\mathrm{N_{TB}}$ order, and gives rise to a net polarization $\mathbf{P}$ that lies principally in the plane perpendicular to the PFC axis. 
\item In a coarse-grained picture, the process is analogous to the electroclinic effect in a chiral smectic-A phase: Tilting of the average director (helical axis) away from the ``pseudo-layer" normal induces a $\mathbf{P}$ orthogonal to the tilt plane (with the direction in this plane depending on the sign of the helicity).
\item The locally broken centrosymmetry, together with the electroclinic analogy, determine the structure of the second-order nonlinear susceptibility tensor $\mathbf{\chi}^{(2)}$.
\item The mirror symmetry of the ``pseudo-layer" displacement through planes containing the PFC axis and a pair of parabolic ``arms" leads to cancellations of $\mathbf{P}$ between opposing ``arms". While this complicates development of a quantitative relation between the local magnitude of $\mathbf{P}$ and the observed SH signal, it still allows an explanation of the main qualitative features (e.g., polarization selection rules and angular distribution) of the SH signal. 
\end{enumerate}

\subsection{Theoretical model for induced polarization}

Fig.~5(a) shows a three-dimensional rendering of deformed layers in the core region of a PFC, which is based upon the geometrical construction given in Ref.~\cite{Rosenblatt_PFC}. Although this construction strictly applies in the limit of incompressible layers (i.e., infinite elastic constant for layer compression), it provides a useful starting point. 

In a normal smectic-A, curvature of the layers implies a splay of the director field, and thus in principle a flexoelectric polarization normal to the layers and in the direction of $\hat{\mathbf{n}} (\mathbf{\nabla} \cdot \hat{\mathbf{n}})$ \cite{Meyer_flexo}. However, as already pointed out, no SHG from such a mechanism was observed due to PFCs in the smectic-A phase of the control sample 8CB. For the layer distortion in a PFC defect, the maximum curvature occurs at the conical ``cusps", where the layers ``pinch" down to points located along the parabolae (dashed lines in Fig.~5) that define the defect. In order to avoid an infinite splay energy, the molecules must either rotate off the layer normal (relaxing the splay) and adopt a more uniform orientation at the cusp, or the vertex of the cusps must relax from a conical tip (infinite layer curvature) to a bowl shape (finite curvature). In the former case, the splay flexoelectric polarization would be significantly reduced, while in the latter case the curvature of the layer approaching a cusp would have the opposite sign of the curvature at the bottom of the bowl (that replaces the sharp tip of the cusp), again tending to cancel the splay flexoelectricity. These effects can explain the absence of a detectable SH signal in the control sample (ordinary smectic-A). 

In the dimers forming the pseudo-layered $\mathrm{N_{TB}}$ phase, the same terminal groups on identical mesogenic units are connected to opposite ends of the flexible spacer; this architecture tends to eliminate an overall longitudinal molecular dipole, and thus negate the conventional splay flexoelectric effect at the molecular level. As will be made clear below, the polarization selection of the SH light observed in our experiment also cannot be explained by conventional splay flexoelectric polarization, which would have its major component along the PFC axis ($\hat{\mathbf{z}}$ in Fig.~5). Moreover, any bend flexoelectricity associated with the equilibrium (heliconical) NTB structure averages out over a length scale much shorter than an optical wavelength.  

\begin{figure*}[t]
\centering
\includegraphics[width=2.0\columnwidth]{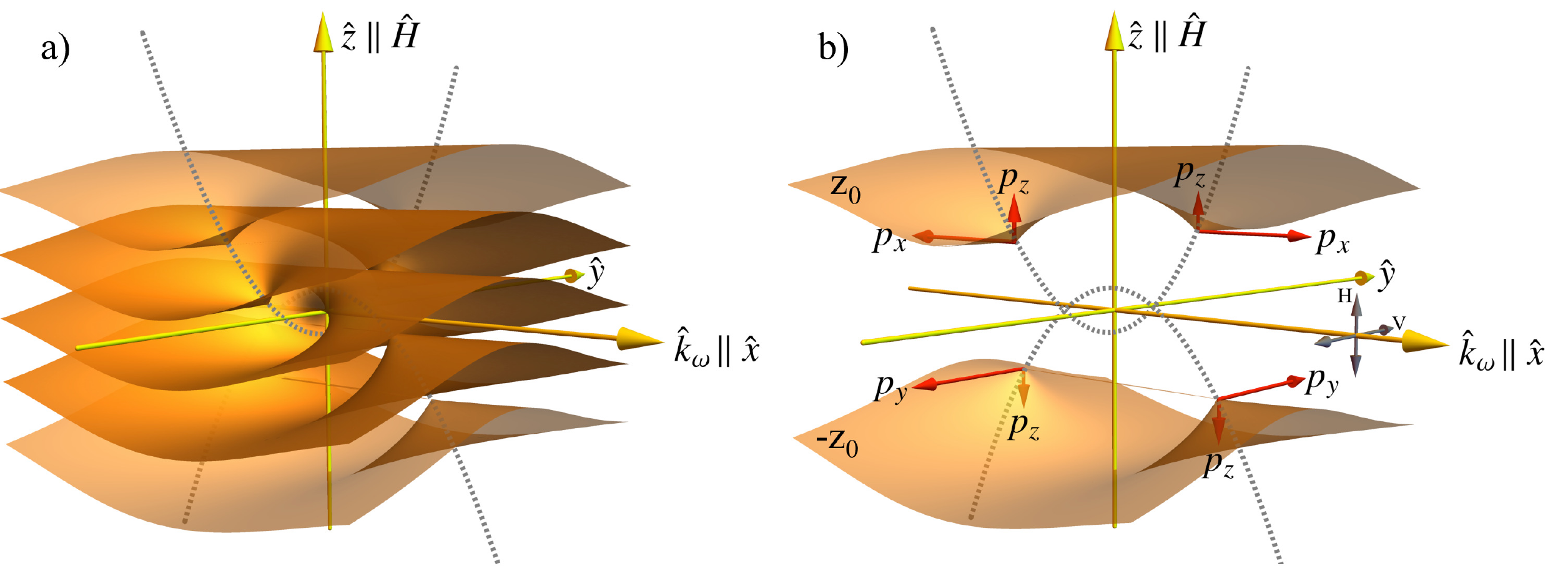}
\caption{(a): 3D rendering of pseudo-layer distortion surrounding the core region of a PFC in the limit of infinite layer compression elastic constant. The two overlapping parabolae (dashed lines) pass through each other's focus and lie in orthogonal planes, whose intersection defines the axis of the PFC. The parabolae are the loci of conical cusps (``pinch" points) in the deformation of the layer planes. Since these are points of infinite curvature energy, they will relax into rounded ``bowls" in the actual case of finite compression constant. The PFC axis corresponds to the direction of average director alignment (applied magnetic field direction $\mathbf{H}$) in our experiment. An orthogonal direction, perpendicular to the plane of one parabola, corresponds to the wavevector of the incident (fundamental light), $\hat{\mathbf{k}}_\omega$ and to the forward direction of the transmitted SH light. (b): Two distorted ``pseudolayers", extracted from the stack in (a) and located at equal distances ($\pm z_0$) from the center ($z=0$) of the PFC. The components of the induced polarization field $\mathbf{p}$ due to an imposed layer displacement $u$ along $\hat{\mathbf{z}}$, as predicted from the model discussed in the text, are shown in the cusp regions (assumed rounded in the limit of finite layer compression constant). Note in particular that the model predicts that the component of $\mathbf{p}$ perpendicular to the PFC axis is substantially larger than that along it. The fundamental and second harmonic polarization axes ($V$ and $H$) are also indicated; these correspond to vertical and horizontal directions with respect to the $\hat{\mathbf{k}}_\omega - \mathbf{H}$ plane of the experiment (and not with respect to those directions along the page.}
\end{figure*}    

Thus, we turn to a model that explicitly couples long wavelength distortions of the heliconical director structure to distortions of a helical vector order parameter $\mathbf{P}$ (where $\mathbf{P}$ may originate either from a net transverse dipole moment, or simply from the ``shape" polarization, associated with the bent conformation of LC dimers. It is convenient to introduce a dimensionless form for this order parameter, $\mathbf{p} = \mathbf{P}/P_{sat}$, where $P_{sat}$ corresponds to the saturated polarization at low temperature.

The $\mathrm{N_{TB}}$ free energy density may then be expanded in terms of the uniaxial director field $\hat{\mathbf{n}}$ and the polarization field $\mathbf{p}$ as \cite{Shamid_PRE}
\begin{eqnarray}
\label{fntb}
F_{NTB} &=& \frac{K_1}{2} (\nabla \cdot \hat{\mathbf{n}})^2 + \frac{K_2}{2} (\hat{\mathbf{n}} \cdot \nabla \times \hat{\mathbf{n}})^2\\
&&+ \frac{K_3}{2} [\hat{\mathbf{n}} \times (\nabla \times \hat{\mathbf{n}})]^2 + \frac{\mu}{2} |\mathbf{p}|^2 + \frac{\nu}{4} |\mathbf{p}|^4 \nonumber\\
&&+ \frac{\kappa}{2} (\nabla \mathbf{p})^2 - \Lambda [\hat{\mathbf{n}} \times (\nabla \times \hat{\mathbf{n}})] \cdot \mathbf{p} + \eta (\hat{\mathbf{n}} \cdot \mathbf{p})^2 . \nonumber
\end{eqnarray}
Here, $K_1$, $K_2$, and $K_3$ are the usual Frank elastic constants for splay, twist, and bend distortions of the nematic director $\hat{\mathbf{n}}$.  The coefficient $\mu = \mu_0 (T - T_0)$ is the temperature-dependent Landau coefficient for the polarization $\mathbf{p}$ ($\mu_0$ being a constant), while $\nu>0$ is a higher-order, temperature-independent Landau coefficient.  The elastic constant $\kappa$ penalizes spatial distortions in $\mathbf{p}$, and the coefficient $\Lambda$ couples $\mathbf{p}$ with bend distortions.  The last term (not included in Ref.~\cite{Shamid_PRE}), with $\eta > 0$, penalizes any component of $\mathbf{p}$ that lies along $\hat{\mathbf{n}}$. Because $\mathbf{p}$ is defined as a dimensionless quantity, the Landau coefficients $\mu$ and $\nu$ carry the same units, and $\kappa$ has the same units as the Frank constants.

In the $\mathrm{N_{TB}}$ phase, the director field has the equilibrium heliconical modulation $\hat{\mathbf{n}} = \hat{\mathbf{z}}\cos\beta+\hat{\mathbf{x}}\sin\beta\cos(q_0 z)+\hat{\mathbf{y}}\sin\beta\sin(q_0 z)$, with pitch wavenumber $q_0$ and cone angle $\beta$. Likewise, the polarization field has the equilibrium helical modulation $\mathbf{p} =\hat{\mathbf{x}}p_0\sin(q_0 z)-\hat{\mathbf{y}}p_0\cos(q_0 z)$, with magnitude $p_0$, perpendicular to $\hat{\mathbf{n}}$ and to the pitch axis $\hat{\mathbf{z}}$, as shown in Fig.~1(b).  In the nematic phase, $\beta$ and $p_0$ are both zero while $q_0$ is undefined; in the $\mathrm{N_{TB}}$ phase, these quantities all become non-zero.

Let us now consider an imposed spatial variation in the phase, $\Phi$, of the helicoidal modulation. This corresponds to a displacement, $u = -\Phi/q_0$, of the $\mathrm{N_{TB}}$ ``pseudo-layers" along the equilibrium layer normal $\hat{\mathbf{z}}$. Under the imposed distortion, the layer plane varies in orientation, and the directions of $\hat{\mathbf{n}}$ and $\mathbf{p}$ also generally change in order to minimize $F_{NTB}$. To describe such a change in $\hat{\mathbf{n}}$, we introduce the orthogonal triad of unit vectors $\hat{\mathbf{e}}_1$, $\hat{\mathbf{e}}_2$, $\hat{\mathbf{t}}$, where $\hat{\mathbf{t}}$ is obtained by rotation of $\hat{\mathbf{z}}$ about an arbitrary axis in the $x$-$y$ plane, and $\hat{\mathbf{e}}_1$ and $\hat{\mathbf{e}}_2$ are the transformed $\hat{\mathbf{x}}$ and $\hat{\mathbf{y}}$ under this rotation. Then the distorted $\hat{\mathbf{n}}$ is
\begin{eqnarray}
\hat{\mathbf{n}} &=& \hat{\mathbf{e}}_1 \sin\beta \cos(q_0 z - q_0 u) + \hat{\mathbf{e}}_2 \sin\beta \sin(q_0 z - q_0 u) + \nonumber\\
&& \hat{\mathbf{t}} \cos\beta.
\end{eqnarray}
The rotated coordinate unit vectors may be expressed in the $xyz$ system as
\begin{eqnarray}
&&\hat{\mathbf{t}} = t_x \hat{\mathbf{x}} + t_y \hat{\mathbf{y}} + t_z \hat{\mathbf{z}},\nonumber\\
&&\hat{\mathbf{e}}_1 = \frac{t_y^2 + t_x^2 t_z}{t_x^2 + t_y^2} \hat{\mathbf{x}} + \frac{t_x t_y (-1+t_z)}{t_x^2+t_y^2} \hat{\mathbf{y}} - t_x \hat{\mathbf{z}},\nonumber\\
&&\hat{\mathbf{e}}_2 = \frac{t_x t_y (-1+t_z)}{t_x^2+t_y^2} \hat{\mathbf{x}} + \frac{t_x^2 + t_y^2 t_z}{t_x^2 + t_y^2} \hat{\mathbf{y}} - t_y \hat{\mathbf{z}},
\end{eqnarray}
where $t_z = \sqrt{1-t_x^2-t_y^2}$, and $t_x$, $t_y$ are the $x$, $y$ components of $\hat{\mathbf{t}}$.

Since $\mathbf{p}$ is not a unit vector and need not, in general, remain orthogonal to $\hat{\mathbf{n}}$ under the imposed distortion, we write,
\begin{eqnarray}
\mathbf{p} &=& \hat{\mathbf{e}}_1 p_0\sin(q_0 z - q_0 u)-\hat{\mathbf{e}}_2 p_0\cos(q_0 z - q_0 u) + \nonumber\\ 
&& + p_x \hat{\mathbf{x}} + p_y \hat{\mathbf{y}} + p_z \hat{\mathbf{z}}.
\end{eqnarray}
Here $p_x$, $p_y$, $p_z$ are the $xyz$ components of the distortion in $\mathbf{p}$ that does not remain perpendicular to $\hat{\mathbf{n}}$.

The calculation of these components for an imposed ``pseudo-layer" displacement $u=u(x)$, which varies along an axis in the equilibrium layer plane (and therefore describes layer bending), and for $u=u(z)$, which varies along the layer normal (describing layer compression), is outlined in the Appendix. For simplicity, the calculation is done only to linear order in $\{p_x,p_y,p_z,t_x,t_y\}$. 

The results, assuming large $\eta$ \cite{Parsouzi_PRX} and small $\beta$ ($\beta \lesssim 10^\circ$ close to the $\mathrm{N_{TB}}$-N transition \cite{Borshch_Nature,CMeyer3}, are: 
\begin{eqnarray}
&&p_x \approx -\frac{\Lambda q_0}{2 \eta} t_y,\\
&&p_y \approx \frac{\Lambda q_0}{2 \eta} \left( t_x + \frac{du}{dx} \right),\\
&&p_z \approx \frac{\Lambda \beta^2}{4 \eta} \left( \frac{dt_x}{dx} - q_0 t_y \frac{du}{dx} \right),
\end{eqnarray}
for $u=u(x)$, and
\begin{eqnarray}
&&\mathbf{p}_\perp \approx -\frac{\sqrt{2} \Lambda}{\eta \beta^2} \frac{d \mathbf{t}_\perp}{dz},\\
&&p_z \approx 0,
\end{eqnarray}
for $u = u(z)$. Here $\mathbf{p}_\perp = p_x \hat{\mathbf{x}} + p_y \hat{\mathbf{y}}$ and similarly for $\mathbf{t}_\perp$. The approximate equality signs in Eqs.~(5)-(9) reflect the approximation associated with our linear analysis in $\{p_x,p_y,p_z,t_x,t_y\}$. 

Now let us apply these results to the cusp region of a PFC observed in the $\mathrm{N_{TB}}$ phase. We employ them with the understanding that they are based on a linear analysis, and thus only represent leading order contributions to $\mathbf{p}$ in the case of ``pseudo-layer" distortion associated with a PFC. We also assume that the size of the cusp region is small compared to the size of a homochiral domain (i.e., domain with single handedness of the underlying heliconical director structure) \cite{CMeyer2}, so that the polarization $\mathbf{p}$ is not washed out in the cusp region due to mixed left and right-handed helical domains.  

As mentioned above and particularly in the case of the $\mathrm{N_{TB}}$ phase, where the ``pseudo-layer" compression elastic constant is substantially lower than that for a typical smectic LC \cite{Challa,CMeyer1}, the sharp tips of the cusps should realistically be envisioned as rounded caps. This also means that in addition to ``pseudo layer" bending, there will be a compression of the layers in the cusp region. However, as described in the Appendix, for $du/dz \ll 1$, the Euler-Lagrange equation for $\mathbf{t}_\perp$ gives $\mathbf{t}_\perp = 0$ to lowest order, and in that case Eqs.~(8) and (9) imply $\mathbf{p} \approx 0$ for ``pseudo-layer" compression. Therefore, we will focus on the effect of layer bending. 

Consider then a pair of distorted layers equidistant from the center of the PFC, at $+z_0$ and $-z_0$, along the $z$ axis, as shown in Fig.~5(b), and consider first the behavior in a slice taken through the cusp in the $x-z$ plane for the layer at $-z_0$. Along $x$, the displacement $u$ is asymmetric through the cusp region and thus $t_x$ will be also; this will lead to a net $p_y$ and $p_z$, according to Eqs.~(6)--(7). On the other hand, for $u=u(x)$, $t_y = 0$ to lowest order (see Appendix), and thus Eq.~(5) implies $p_x \approx 0$ in the $-z_0$ layer. On opposite sides of the $x$ axis, i.e., $x \rightarrow -x$, the direction of $p_y$ is reversed. 

Next, for a slice taken in the $y-z$ plane, we may interchange $x$ and $y$ in Eqs.~(5)--(7); however, the mirror symmetry across this plane means that only the $p_z$ component of the polarization will survive. The net components of $\mathbf{p}$ in the cusp region are shown qualitatively in Fig.~5(b); the relative magnitudes shown for $p_y$ and $p_z$ will be explained shortly. 

For the layer at $+z_0$, the results are the same as for $-z_0$, except for a $90^\circ$ rotation around the $z$ axis, equivalent to an exchange of $p_x$ and $p_y$ and a reversal in the direction of $p_z$ (see Fig.~5(b)).

The relative magnitudes of $p_y$ and $p_z$ in the $-z_0$ layer (or $p_x$ and $p_z$ in the $+z_0$ layer) may be estimated if we consider a harmonic form for $u$ -- specifically, a Fourier component whose wavenumber $q_x = 2 \pi/\delta$ corresponds to the length scale of the cusp region of the PFC. That scale can be estimated \cite{Rosenblatt_PFC} from the width $\delta \simeq 2 \mu$m of the birefringent arms of the parabolae (the loci of PFC cusps) in the textures presented in Fig.~2. Then taking $u = u(x) = u_0 \exp(i q_x x)$, the Euler Lagrange equation for $t_x$ yields (see Appendix)
\begin{equation}
t_x \approx -\frac{K_3 q_0^2 \beta^2}{K_1 q_x^2 + K_3 q_0^2 \beta^2} \frac{du}{dx}.
\end{equation}
Typically, $\beta = 0.18$~rad, $q_0 = 2 \pi / (0.01 \mu \mathrm{m})$, and $q_x = 2 \pi/(2 \mu \mathrm{m})$, so $q_x^2 \ll q_0^2 \beta^2$. Then the right hand side of the expression for $t_x$ may be expanded, yielding
\begin{equation}
t_x \approx \left( -1 + \frac{K_1 q_x^2}{K_3 q_0^2 \beta^2} \right) \frac{du}{dx}
\end{equation} 
Inserting this result into Eqs.~(6) and (7) with $u = u_0 \exp(i q_x x)$, we obtain to lowest order 
\begin{eqnarray}
p_y &\approx& i \frac{\Lambda}{2 \eta} \frac{K_1}{K_3} \frac{q_x^3}{q_0 \beta^2} u,\\
p_z &\approx& \frac{\Lambda}{4 \eta} \beta^2 q_x^2 u.
\end{eqnarray}
Then, using the numerical values given above, and taking $K_1 \simeq 2 K_3$ (for the bare nematic values) \cite{Borshch_Nature}, we estimate 
\begin{equation}
\left| \frac{p_z}{p_y} \right| \approx \frac{\beta^4}{2} \frac{K_3}{K_1} \frac{q_0}{q_x} \approx 0.05
\end{equation}

Since the SH signal should scale as the square of $\mathbf{p}$ (assuming the components of the second order susceptibility $\mathbf{\chi}^{(2)}$ scale with the magnitude of $\mathbf{p}$), the contribution from $p_z$ is negligible. A similar result applies to the layer at $+z_0$. In the following subsection, we will therefore concentrate only on the component of $\mathbf{p}$ perpendicular to the axis of the PFC.  

\subsection{Nonlinear susceptibility in the $\mathrm{N_{TB}}$ phase}

Although SHG could possibly arise due to chiral symmetry breaking in the undistorted $\mathrm{N_{TB}}$ state, this structural chirality averages out over a molecular length scale (the heliconical pitch) that is much shorter than the optical wavelength. We detected SH signal from our samples only in the presence of pseudo-layer distortion specifically associated with PFCs. The situation is similar to the chiral smectic-C* phase, where the helical polarization must be partly unwound by an applied field in order to observe a signal \cite{FLC_NLO}. 

Therefore, we propose that the broken centrosymmetry produced by the induced polarization $\mathbf{p}$, over length scales comparable to or longer than the optical scale, is responsible for the observed SHG. As shown in Fig.~5(b), and based on the arguments above, the dominant component of the distortion-induced polarization is normal to the axis of the PFC and to the direction of the applied magnetic field, both of which are along $\hat{\mathbf{z}}$. Eq.~(6)and the discussion above reveal that the polarization arises essentially from an electroclinic effect, where the coarse-grain director tilts with respect to the pseudo-layer normal and the polarization develops along the axis perpendicular to the tilt plane. Indeed, for small distortions, the tilt angle is given by the sum $t_x + du/dx$ in Eq.~(6). 

The local symmetry is therefore that of a tilted chiral smectic -- namely, $C_2$ symmetry, with symmetry axis ($\mathbf{p}$) perpendicular to the axis of the PFC ($\hat{\mathbf{z}}$). As evidenced in Fig.~2, the plane of one parabola of the PFCs is oriented parallel to the substrates. We take this as the $y-z$ plane and the incident (fundamental) light propagation direction as $\hat{\mathbf{k}}_\omega =\hat{\mathbf{x}}$. Then the polarizations $V$ or $H$ of the fundamental or second harmonic waves are $V = \hat{\mathbf{y}}$ and $H = \hat{\mathbf{z}}$, and the second harmonic intensities for the various polarization combinations are,
\begin{eqnarray}
&&I (H_{\omega} \rightarrow V_{2 \omega}) \propto |\chi_{VHH}^{(2)}|^2 = |\chi_{yzz}^{(2)}|^2, \nonumber\\
&&I (H_{\omega} \rightarrow H_{2 \omega}) \propto |\chi_{HHH}^{(2)}|^2 = |\chi_{zzz}^{(2)}|^2, \nonumber\\
&&I (V_{\omega} \rightarrow H_{2 \omega}) \propto |\chi_{HVV}^{(2)}|^2 = |\chi_{zyy}^{(2)}|^2, \nonumber\\
&&I (V_{\omega} \rightarrow V_{2 \omega}) \propto |\chi_{VVV}^{(2)}|^2 = |\chi_{yyy}^{(2)}|^2.
\end{eqnarray} 
Here $\mathbf{\chi}^{(2)}$ is the second order susceptibility for $C_2$ symmetry, with symmetry axis = $\hat{\mathbf{y}}$ ($-z_0$ layer in Fig.~5(b)) or $\hat{\mathbf{x}}$ ($+z_0$ layer). Assuming Kleinman's symmetry, the non-vanishing components of $\mathbf{\chi}^{(2)}$ with recurring indices are \cite{Boyd_NLO},
\begin{equation}
\chi_{xxy}^{(2)} = \chi_{xyx}^{(2)} = \chi_{yxx}^{(2)}~,~\chi_{zzy}^{(2)} = \chi_{zyz}^{(2)} = \chi_{yzz}^{(2)}~,~\chi_{yyy}^{(2)} \nonumber
\end{equation}
for $-z_0$ layers, and
\begin{equation}
\chi_{yyx}^{(2)} = \chi_{yxy}^{(2)} = \chi_{xyy}^{(2)}~,~\chi_{zzx}^{(2)} = \chi_{zxz}^{(2)} = \chi_{xzz}^{(2)}~,~\chi_{xxx}^{(2)} \nonumber
\end{equation}
for $+z_0$ layers. From these lists and the relations for SH intensity in Eq.~(15), we conclude that the intensity for $H$ polarized output vanishes, in agreement with the polarization selection observed in our experiment, and that the SH signal is generated predominantly by the induced polarization in the $-z_0$ layers (where $\mathbf{p} \parallel \hat{\mathbf{y}}$) in Fig.~5(b).

Fig.~3 reveals that the SH intensity is systematically lower for a $V_{\omega} \rightarrow V_{2\omega}$ process than for $H_{\omega} \rightarrow V_{2\omega}$, indicating $|\chi_{yzz}^{(2)}|^2 > |\chi_{yyy}^{(2)}|^2$ according to Eq.~(15). We can then suggest the following explanation for the splitting of the forward peak in the $V_{\omega} \rightarrow V_{2\omega}$ data: In linear light scattering, we have observed strong depolarized scattering from defect structure at small angles in the $\mathrm{N_{TB}}$ phase. Consider then a process by which small-angle linear scattering converts polarization $V_{\omega}$ to $H_{\omega}$, and then a SH process that converts $H_{\omega}$ to $V_{2\omega}$. This could lead to a slight splitting of the forward peak, as observed in Fig.~3 for the $V_{\omega} \rightarrow V_{2\omega}$ data. On the other hand, the sequence $H_{\omega} \rightarrow V_{\omega} \rightarrow V_{2\omega}$ would be less efficient, producing perhaps a broadening but not a splitting of the peak in the $H_{\omega} \rightarrow V_{2\omega}$ data. Moreover, the peak for $V_{\omega} \rightarrow V_{2 \omega}$ splits along the magnetic-field alignment direction -- i.e., parallel to the axis of the PFCs. This is because the distortion-induced polarization near the core region will be concentrated more along the PFC axis than normal it.         

Finally, we note that the major components of the induced polarization are oppositely directed on opposite sides of the parabola axis (Fig.~5(b)), even if the entire PFC is in a homochiral domain. In this case, there will be no net SH output when the distance across the parabola yields a destructive interference condition, or when the vector sum of polarizations cancels out over a length scale small compared to the coherence length $\lambda_{2 \omega}/(n_{2 \omega} - n_\omega) \simeq 5~\mu$m ($n =$~refractive index) \cite{Boyd_NLO}. As mentioned in the Results section, this effect accounts for the relatively weak signal recorded in the experiment. The phase factor is important, and consequently the signal level is not a direct measure of the {\it amplitude} of the helical polarization field. 

\section{Conclusion}

In this paper, we have used second harmonic light scattering to probe polar molecular organization in the nematic twist-bend phase. The SH signal arises from parabolic focal conic defects in the ``pseudo-layer" structure of the $\mathrm{N_{TB}}$ state. Its key features can be explained by a coarse-grained free energy that couples the ``layer" deformation to long wavelength distortions of a helical polarization wave, which serves as the $\mathrm{N_{TB}}$ order parameter. In the future, we hope to employ focused fundamental radiation in order to generate SH light from specific regions of an individual PFC defect in a thin sample. This approach, combined with a more detailed treatment of the PFC structure, could provide a direct measure of the magnitude of the $\mathrm{N_{TB}}$ polarization field.

\begin{acknowledgments}
We are indebted to our colleague Prof. O. D. Lavrentovich for several informative conversations, and to Dr. Owain Parri for providing us with the material KA(0.2) used in this study. This work was supported by the NSF under grants DMR-1307674 (SP, JG, AJ, SS) and DMR-1409658 (JVS); the EPSRC under grant EP/J004480/1 (GM and CW); and the EU under project 216025 (MGT).
\end{acknowledgments}

\section*{Appendix: Induced polarization due to ``pseudo-layer" distortion}
In this appendix, we sketch the details leading to Eqs.~(5)--(10) of the text. We first consider a ``pseudo-layer" displacement $u$ that varies along a single direction in the plane of the equilibrium ``pseudo-layers" -- e.g., $u = u(x)$. This corresponds to a bending of the layers without compression along the average layer normal ($\hat{\mathbf{z}}$). 

Expressions for the induced $\{p_x,p_y,p_z,t_x,t_y\}$ that minimize the free energy under the imposed $u(x)$ may be obtained by the following procedure: (1) Combine Eqs.~(2)--(4) in the text and insert the result into the free energy density, Eq.~(1); (2) ``Coarse-grain" the free energy density by averaging with respect to $z$ over a helicoidal pitch, $2 \pi / q_0$, based on the assumption that $u$ and the variables $\{p_x,p_y,p_z,t_x,t_y\}$ vary slowly over the length scales comparable to the pitch; (3) Obtain and solve the Euler-Lagrange equations for the set $\{p_x,p_y,p_z,t_x,t_y\}$ that minimize the coarse-grained free enrgy density. In step (2), the coarse-grained $F$ is evaluated as $F_{cg} (\hat{\mathbf{t}},u,\mathbf{p};x) = \frac{q_0}{2 \pi} \int_{0}^{2 \pi/q_0} F (\hat{\mathbf{n}},\mathbf{p};x,z) \, dz$. The assumption on relative length scales is reasonable, since the pitch is on the molecular scale, while the features associated with the PFC textures observed in the $\mathrm{N_{TB}}$ phase have sizes $\gtrsim 1 \mu$m. 

After implementing this program, and assuming that terms containing the elastic constant $\kappa$ are neglgible (i.e., long wavelength limit for the distortion $u(x)$ and small polarization elastic constant $\kappa$ \cite{Parsouzi_PRX}, we find to lowest order in $\{p_x,p_y,p_z,t_x,t_y\}$,
\begin{eqnarray}
p_x &=& - \frac{\Lambda q_0 \sin^2 \beta}{2 (\mu + 2 \nu p_0^2 + \eta \sin^2 \beta)} t_y ,\\
p_y &=& \frac{\Lambda q_0 \sin^2 \beta}{2 (\mu + 2 \nu p_0^2 + \eta \sin^2 \beta)} \left( t_x + \frac{du}{dx} \right) ,\\
p_z &=& \frac{\Lambda \sin^2 \beta}{2 (2\eta \cos^2\beta + \mu + \nu p_0^2)} \times \nonumber\\
&&\left( \frac{dt_x}{dx} - q_0 t_y \frac{du}{dx} \right).
\end{eqnarray}
For $u = u(x)$, $\{p_x,p_y,p_z,t_x,t_y\}$ are functions of $x$. Taking the small $\beta$ limit of Eqs.~(A1)-(A3) yields Eqs.~(5)-(7) of the text.

To linear order in $\{t_x,t_y\}$, the Euler-Lagrange equations that must be solved to obtain $\hat{\mathbf{t}}$ in terms of a given $u(x)$, and thus complete the solution for $\mathbf{p}$, are
\begin{eqnarray}
&&-\frac{A_x}{2} \frac{d^2t_x}{dx^2} + \frac{B_x}{2} \left( t_x + \frac{du}{dx} \right) = 0,\\
&&-\frac{A_y}{2} \frac{d^2t_y}{dx^2} + \frac{B_y}{2} t_y = 0,
\end{eqnarray}
where the coefficients (together with their small $\beta$ limits) are given by
\begin{eqnarray}
A_x &=& K_1 (\cos2\beta+1) + \frac{K_2}{32} (3-4\cos2\beta+\cos4\beta) + \nonumber\\
&&\frac{K_3}{32} (28+4\cos2\beta)\sin^2\beta - \nonumber\\
&&\frac{\Lambda^2 \sin^4\beta}{2(2 \eta \cos^2\beta+\mu+\nu p_0^2)} \nonumber\\
&\approx& 2 K_1,\\
B_x &=& B_y = \Lambda p_0 q_0 \sin2\beta +\frac{q_0^2 \sin^2\beta}{2} \biggl[ 2 K_1 + \nonumber\\
&&(3\cos2\beta-1)K_2 - \frac{\Lambda^2 \sin^2\beta}{2(\mu+2\nu p_0^2+\eta \sin^2\beta)} \biggr] \nonumber\\
&\approx& 2 \Lambda p_0 q_0 \beta, \\
A_y &=& \frac{41+20\cos2\beta+3\cos4\beta}{32} K_2 + \nonumber\\
&& \frac{7 - 4\cos2\beta - 3 \cos4\beta}{32} K_3 \nonumber\\
&\approx& 2 K_2.
\end{eqnarray}

Note that Eq.~(20) is independent of the imposed distortion $u$. Thus, to linear order the only physically acceptable solution is $t_y = 0$, implying $p_x = 0$ in Eq.~(16).

Next we consider a ``pseudo-layer" displacement that varies along the axis normal to the equilibrium layer planes -- i.e., a layer compression, with $u = u(z)$. Carrying through a similar analysis as above, we find to lowest order in $\{\mathbf{p}_\perp = p_x \hat{\mathbf{x}} + p_y \hat{\mathbf{y}}~,~\mathbf{t}_\perp = t_x \hat{\mathbf{x}} + t_y \hat{\mathbf{y}}\}$, and neglecting terms containing $\kappa$,
\begin{eqnarray}
&&\mathbf{p}_\perp = -\frac{\sqrt{2} \Lambda \cos^2 \beta}{\mu + 2 \nu p_0^2 + \eta \sin^2 \beta} \frac{d\mathbf{t}_\perp}{dz},\\
&&p_z = 0,
\end{eqnarray}
Here $\{p_\perp, t_\perp\}$ depend on $z$. Eq.~(8) in the text corresponds to the large $\eta \sin^2 \beta$ and small $\beta$ limit of Eq.~(24).

The Euler-Lagrange equation for $\mathbf{t}_\perp$ (linearized in $\mathbf{t}_\perp$) is
\begin{equation}
-\frac{A_\perp}{2} \frac{d^2 \mathbf{t}_\perp}{dz^2} + \left( B_\perp + C_\perp \frac{du}{dz} \right) \left( 1 - \frac{du}{dz} \right) \mathbf{t}_\perp = 0,
\end{equation}
where
\begin{eqnarray}
A_\perp &=& K_1 + \frac{K_2}{4} + \frac{7K_3}{4} + (2K_3-K_1) \times \nonumber\\
&&\cos2\beta + \frac{(K_3-K_2) \cos4\beta}{4} + \nonumber\\
&&\frac{2 \Lambda^2 \cos^4\beta}{\mu + 2 \nu p_0^2 + \eta \sin^2\beta},\\
B_\perp &=& C_\perp + \Lambda p_0 q_0 \sin2\beta, \\
C_\perp &=& -2q_0^2\sin^2\beta \biggl[ K_1+ \frac{3\cos2\beta-1}{2}  \times \nonumber\\
&&(K_2-K_3) \biggr].
\end{eqnarray}
For $du/dz \ll 1$, Eq.~(26) reduces to $-(A_\perp/2) d^2\mathbf{t}_\perp/dz^2 + B_\perp \mathbf{t}_\perp = 0$, the only physical solution to which is $\mathbf{t}_\perp = 0$.

Finally, if we consider a simple harmonic form for $u(x)$, $u(x) = u_0 \exp (i q_x x)$, then Eqs.~(19), (21) and (22) yield (for small $\beta$),
\begin{equation}
t_x = -\frac{p_0 q_0 \Lambda \beta}{K_1 q_x^2 + p_0 q_0 \Lambda \beta} \frac{du}{dx}
\end{equation}
Since for small $\beta$ the polarization magnitude $p_0$ in the $\mathrm{N_{TB}}$ phase is \cite{Parsouzi_PRX} $p_0 \approx K_3 q_0 \beta/\Lambda$, Eq.~(30) produces Eq.~(10) of the text.

\end{document}